\documentclass[11pt]{article} 

\usepackage[utf8]{inputenc} 

\usepackage{geometry} 
\usepackage{graphicx} 

\usepackage{booktabs} 
\usepackage{array} 
\usepackage{paralist} 
\usepackage{verbatim} 
\usepackage{subfig} 
\usepackage{amsmath} 
\usepackage{amssymb} 
\usepackage{epstopdf} 
\usepackage{color}

\usepackage{fancyhdr} 
\usepackage{sectsty}
\allsectionsfont{\sffamily\mdseries\upshape} 

\usepackage[nottoc,notlof,notlot]{tocbibind} 
\usepackage[titles,subfigure]{tocloft} 

\title{Threshold-limited spreading in social networks with multiple initiators}

\author{P. Singh$^{1,2~}\footnote{E-mail: singhp4@rpi.edu}$, S. Sreenivasan$^{1,2,3}$, B.K. Szymanski$^{2,3}$, G. Korniss$^{1,2}$}

\begin{document}
\maketitle

\begin{flushleft}
$^{\bf{1}}$ Department of Physics, Applied Physics, and Astronomy, Rensselaer Polytechnic Institute,
110 8$^{th}$ Street, Troy, NY, 12180-3590 USA \\
$^{\bf{2}}$ Social Cognitive Networks Academic Research Center,
Rensselaer Polytechnic Institute, 110 8$^{th}$ Street, Troy, NY, 12180-3590 USA \\
$^{\bf{3}}$ Department of Computer Science,
Rensselaer Polytechnic Institute, 110 8$^{th}$ Street, Troy, NY, 12180-3590 USA \\
\end{flushleft}

\section*{Abstract}
A classical model for social-influence-driven opinion change is the
threshold model. Here we study cascades of opinion change driven by
threshold model dynamics in the case where multiple {\it initiators}
trigger the cascade, and where all nodes possess the same adoption
threshold $\phi$. Specifically, using empirical and stylized models
of social networks, we study cascade size as a function of the
initiator fraction $p$. We find that even for arbitrarily high value
of $\phi$, there exists a critical initiator fraction $p_c(\phi)$
beyond which the cascade becomes global. Network structure, in
particular clustering, plays a significant role in this scenario.
Similarly to the case of single-node or single-clique initiators
studied previously, we observe that community structure within the
network facilitates opinion spread to a larger extent than a
homogeneous random network. Finally, we study the efficacy of
different initiator selection strategies on the size of the cascade
and the cascade window.


\section*{Introduction}

It has long been known through empirical studies that in a
population of socially interacting individuals where each individual
node holds an opinion from a binary set, a small fraction of {\it
initiators} holding opinion opposite to the one held by the majority
can trigger large cascades and eventually result in a dominant
majority holding the initiators' opinion. Some recent studies have
investigated such phenomena in the context of the adoption of
scientific, cultural, and commercial products
\cite{Uzzi_buzz,Makse_PNAS2013}. One of the simplest models that
captures adoption dynamics, irrespective of context, is the
threshold model \cite{granovetter,Watts02,WattsDodds,centola}.
According to the threshold model, an individual changes its opinion
only if a critical fraction of its neighbors have already adopted
the new opinion. This required fraction of new adoptees in the
neighborhood is designated the {\it adoption threshold}
\cite{granovetter,Latane_JPSP1996}. Here, we denote the adoption
threshold by $\phi$. Since its introduction ~\cite{granovetter}, the
threshold model has been studied extensively on complex networks to
analyze the conditions under which a  vanishingly small fraction (of
the total system size) of initiators is capable of triggering a
cascade of opinion change \cite{Watts02,centola,Qiming_RGG}. In
particular, these studies considered initial conditions with a
single ``active" node \cite{Watts02} or an active connected clique
(a single node and all of its neighbors) \cite{centola} as
initiators. In this scenario, the condition for global cascades in
connected sparse random networks is $\phi<1/\langle k \rangle$
\cite{Watts02,centola,Qiming_RGG}, where $\langle k\rangle$ is the
average degree of the network.
However, with a few exceptions \cite{Kempe_SIGKDD2003,Chen_IEEE2010,Gleeson_PRE2007}, little
attention has been paid to the question of how the size and the
selection of this initiator fraction affects the spreading of an
opinion in the network, in particular, in the regime where a single
active node or a small clique is insufficient to trigger global cascades.

In case of multiple initiators, how to select these initiators from
among the nodes of the network so as to maximize the spread (cascade
size), remains an open question. To address this issue we compare
three different heuristic ways of selecting a set of initiators with
predefined size, on Erd\H{o}s-R\'enyi (ER) random networks
\cite{ER_graph}. Specifically, we look at the size of the spread for
a varying range of the average degree $\langle k\rangle$ of the ER
networks. As found earlier for the case of cascades triggered by
single initiators~\cite{Watts02,Qiming_RGG}, we find that when the
average degree is too low or too high, large cascades are not
triggered. However, within an intermediate range of $\langle
k\rangle$, large cascades are realized. This range is referred to as
the {\it cascade window} \cite{WattsDodds}. We find that the width
of this cascade window is the largest when the initiator nodes are
selected successively in descending order of degree starting with
the node having the largest degree. We also find that the total time
taken for the cascade to terminate is shortest for this selection
strategy.

In both ER \cite{Watts02} and empirical \cite{centola} networks it
was observed that for a given $\langle k\rangle$, there is a
critical threshold $\phi_{c}$ such that cascades are only triggered
if $\phi < \phi_{c}$ for a single-node or a very small initiator set
~\cite{Watts02,centola}. Here, we systematically study the effect of
varying the initiator fraction $p$ with $\phi$ held fixed, for the
entire range of values of the adoption threshold $\phi$. We find
that for any given threshold $\phi < 1$ there exists a critical
value of the fraction of initiators $p_{c}$, above which global
cascades can be triggered. We discuss the dependence of $p_{c}$ on
$\phi$ which turns out to be a smooth curve separating the two
phases, one in which cascades are observed and the other where
cascades cannot be triggered. This finding constitutes an important
insight into how {\em local} neighborhood-level thresholds can
constrain the emergence of tipping points for cascades on global
scales on sparse graphs. We note that in
Refs.~\cite{Kempe_SIGKDD2003,Chen_IEEE2010}, the authors went beyond
basic heuristic selections of the initiators (targets) by employing
a systematic greedy selection and a scalable influence maximization
algorithm, respectively; however they did not explore the region for
global [$O(N)$] cascades (and the corresponding tipping point
$p_{c}$ of initiators required to trigger them), but rather, only
focused on the $p\ll1$ regime.

In Ref.~\cite{Gleeson_PRE2007}, assuming locally
tree-like structures, the authors developed an asymptotic approach
to approximate the size of the cascades. This method is expected to
work better for random graphs with small average degree (with
negligible presence of triads) and to gradually break down for
graphs with higher $\langle k\rangle$. We will comment on its
applicability in determining the tipping point $p_{c}(\phi)$ in the
Results section.

Details of the network structure beyond the average degree $\langle
k\rangle$, also play an important role in the spreading
process~\cite{ws98}.
The network's degree distribution and the
presence of community structure and local clustering can
significantly affect the dynamics of spreading and vulnerability to
cascades in both social networks (driven by influencing) \cite{centola,qiminglu,Ikeda_JPCS2010} and
infrastructure networks (driven by load-based failures) \cite{Huang_PRE2008}.

To elucidate the effect of clustering, we study the effect of network rewiring on the
cascades triggered by different methods. Specifically, starting from
an empirical network with a community structure and relatively high
clustering, we redistribute the links in the network while
preserving the original degree sequence, using a number of different
methods. The cascade sizes are found to be larger and more likely in
the original network which, in addition to having an inherent
community structure, has much higher clustering coefficient
(essentially capturing the density of triads) \cite{ws98}. These
results indicate that local clustering, just like in the case of a
single node (or single-clique) initiator
\cite{centola,Ikeda_JPCS2010}, facilitates the spreading of global
cascades in the case of multiple initiators as well.

A recent study \cite{Brummitt_PRE2012} also considered
cascades in the threshold model in multiplex networks
(a natural framework and terminology for interdependent networks
\cite{Buldyrev_Nature2010,Gao_NatPhys2012,Brummitt_PNAS2012} in the
social setting). In this case, individuals can be connected by
multiple types of edges (representing multiple kinds of social ties,
e.g., colleagues, friends, or family). It was shown that multiplex
networks facilitate cascades, i.e., increase the social network's
vulnerability to spreading \cite{Brummitt_PRE2012}.

\section*{Results}
In the threshold model, every node in the network can be in one of
the two possible states, $0$ (inactive) or $1$ (active), that can be
also be thought of as signifying distinct binary opinions on an
issue. The typical initial condition for studying threshold model
dynamics is one where all nodes except a minority - the initiators -
are in state $0$. Then, the dynamics proceeds as follows. At each
time step, a node is selected at random. If the node is inactive, it
becomes active if at least a threshold fraction $\phi$ of its
neighboring nodes are active i.e. in state $1$. The active state is
assumed to be permanent i.e. once a node becomes active it remains
active indefinitely. The system evolves according to these rules
until no further activations can occur. The threshold $\phi$, in
general, can be different for every node but for simplicity, we
consider the case where every node has the same threshold. The size
of the cascade at any point during its evolution or after it has
terminated, is quantified by the fraction of active nodes in the
network. In the following sections we discuss the simulation of this
dynamics for various network topologies.

\subsection*{Selection strategies}
The decision that a node will adopt $1$ depends only on the states
of its neighbors. If the fraction of its neighborhood in state $1$
exceeds $\phi$ then the node updates its state. As a result of this
{\it threshold condition} a node's degree plays an important role in
determining how easily it can be influenced. The threshold condition
is more easily satisfied for a low-degree node than a high-degree
node, since the former requires fewer active nodes to be present
than the latter, given a fixed adoption threshold $\phi$ for all
nodes. Similarly, the average degree of the network $\langle
k\rangle$ determines to what extent, if at all, the entire network
can be influenced. For a fixed number of initiators, high degree
nodes are less likely to get influenced because it is more difficult
for their neighbors to satisfy the threshold condition. A high
$\langle k\rangle$ is therefore not a desirable condition for
cascades. On the other hand, for low $\langle k\rangle$, the network
consists of disconnected clusters of sizes less than $O(N)$, and
cascades remain confined to one or few of these clusters.  As a
result, global cascades only become possible in an intermediate
range of $\langle k\rangle$ - the cascade window. In general,
cascade window sizes depend on both, the threshold $\phi$, and the
initiator fraction $p$.

The precise choice of initiators also plays an important role in the
size of the cascade and consequently the cascade window itself. A
strategic selection of initiators can dramatically increase the
average size of the spread, which we denote by $S$. Here, we compare
three heuristic strategies for selecting a set of initiators
constituting a fraction $p$ of the total network size: ({\it i}) random
selection, ({\it ii}) selecting nodes in the descending order of their
degrees, and ({\it iii}) selection in the descending order of $k$-shell
index \cite{kitsak}. In ({\it ii}) and ({\it iii}), the choice of
initiators may not be unique. If there are many sets of initiators
that can be selected for the same degree (or $k$-shell), one of these
sets is selected at random.

\begin{figure}[H!]
\centerline{\includegraphics*[width=160mm]{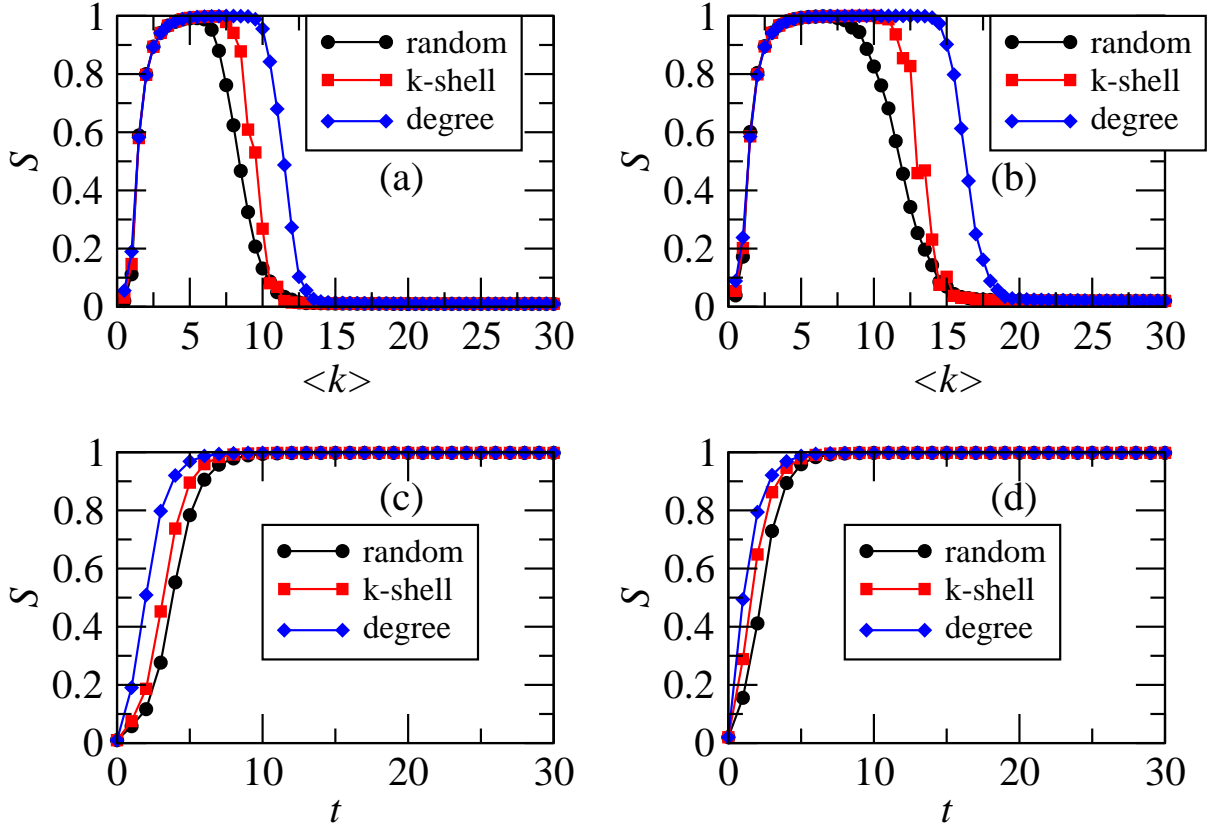}}
\caption{Cascade size $S$ as a function of the average degree on ER
networks of $N=1000$ nodes with threshold $\phi=0.18$ for different
selection strategies of multiple initiators for (a) $p=0.01$; for
(b) $p=0.02$.
Time evolution of the average cascade size $S$ on ER networks of $N=1000$
nodes with average degree $\langle k\rangle=6.0$ and threshold
$\phi=0.18$ for different selection strategies of multiple
initiators for (c) $p=0.01$; for (d) $p=0.02$.}
\label{cwindow}
\end{figure}

The simulation results are shown in Fig. \ref{cwindow}(a) for a
fixed fraction of initiators $p=0.01$ on an ER graph with $N=1000$
and $\phi = 0.18$. We first look at the average spread size as a
function of average degree $\langle k \rangle$ on an ER random graph
as shown in Figure \ref{cwindow}(a). When $\langle k\rangle$ is
small, all three strategies perform equally well because the network
consists only of small clusters without a giant component and hence
spread is localized to those clusters. As soon as $\langle k
\rangle$ becomes large enough for a giant component to arise, the
spread covers a large portion of the network. Further increasing
$\langle k \rangle$ makes it harder for the nodes to satisfy the
threshold condition and $S$ decreases again.

To understand the differences in the performance of these
heuristics, we first note that there are two distinct aspects
determining the efficacy of a node as an initiator. First, it must
be capable of influencing a large number of nodes, i.e. it should
have a large degree. Second, it must be connected to nodes which
have an easily satisfiable threshold condition i.e. the degrees of
its neighbors must be sufficiently low. Additionally, and related to
the first point, it also makes sense to choose the highest-degree
nodes as initiators, since they are the hardest to influence. In
light of these arguments, the highest-degree selection strategy
appears to be a natural choice for generating large cascades. It
would appear that high $k$-shell nodes are a comparably good choice,
since high $k$-shell nodes also possess a high degree. However, by
construction, nodes in the highest $k$-shells are a special subset
of the high-degree nodes that are predominantly connected to other
nodes of high-degree. In other words, nodes selected in descending
order of their $k$-shell index have fewer easily influencable
neighbors than nodes selected purely on the basis of degree. This
qualitatively explains why the $k$-shell method does not perform as
well as the high-degree selection. Finally, the random selection
works the poorest since it largely selects low-degree nodes which
trigger a small number of cascades many of which frequently
terminate when they encounter a high-degree node.

An increase in the initiator fraction $p$ makes the cascade window
wider by allowing cascades to occur for even higher $\langle k
\rangle$ values as shown in Fig.~\ref{cwindow}(b) where $p$ is
increased to $0.02$. The selection strategies follow the same
ranking in this case as well.

Results obtained from simulations indicate that highest degree
method also works better (followed by the $k$-shell method) in terms
of the speed of the cascade. The results for $p=0.01$ and $p=0.02$
are shown in Figs.~\ref{cwindow}(c) and (d), respectively.

\subsection*{Tipping point for multiple initiators}

\begin{figure}[H!]
\centerline{\includegraphics*[width=140mm]{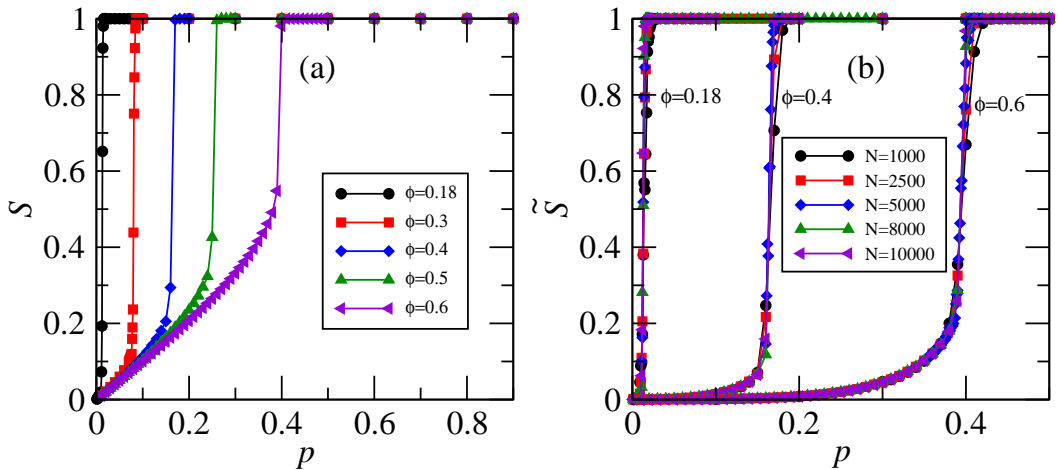}}
\caption{Cascade size and scaled cascade size as a function of initiators on ER networks with $\langle k\rangle=10.0$.
(a) Cascade size $S$ as a function of initiators $p$ for ER networks with $N=10000$ for different values of $\phi$.
(b) Scaled cascade size $\tilde{S}$ [Eq.~(\ref{S_scaled})] vs. $p$ for ER networks with different network sizes $N$ and $\phi$ values.}
\label{sn_p_differentphis}
\end{figure}

As discussed in the previous section, for a small ($O(1)$-size) seed
of initiators, cascades can only occur if $\phi$ is smaller than a
critical value ($\phi<1/\langle k \rangle$ for sparse random graphs
\cite{Watts02,centola,Qiming_RGG}). However, this does not hold if
we introduce a sufficiently large fraction of initiators in the
system. We look at the quantity $S$ (average fraction of nodes in
state $1$) as a function of $p$. (We will refer to $S$ as cascade
size for short.) Gradually increasing $p$ shows that in the
beginning when $p \ll 1$, (global) cascades are not observed. When
$p$ reaches a critical value $p_{c}$, a discontinuous transition
occurs and large cascades are seen immediately as shown in
Fig.~\ref{sn_p_differentphis}(a). The need for a minimum critical
fraction of committed nodes for consensus has been observed in
different models of influence ~\cite{Galam_2007,xie2011,singh2012}
(see Discussion for more details).

Since starting with a finite $p$ itself accounts for a large number
of nodes in state $1$, the relevant quantity to look at is the
number of nodes that were initially in state $0$ and eventually
adopted state $1$ (i.e., excluding the initiators). Thus, we define
\begin{equation}
\tilde{S} = \frac{S-p}{1-p} \;,
\label{S_scaled}
\end{equation}
which measures the fraction of non-initiator nodes that participate
in the cascade. Transitions in $\tilde{S}$ are shown in
Fig.~\ref{sn_p_differentphis}(b) for different $\phi$ values and
several network sizes. It can clearly be seen that the transition
only depends upon $\phi$ and is independent of system size $N$. This
transition (the emergence of the tipping point) is quite generic in
the threshold model, and can be observed in networks with different
sizes and average degrees, as well as for different selection
methods for initiators (see Supplementary Information Sections~S.1
and S.2 for more details).

\begin{figure}[H!]
\centerline{\includegraphics*[width=140mm]{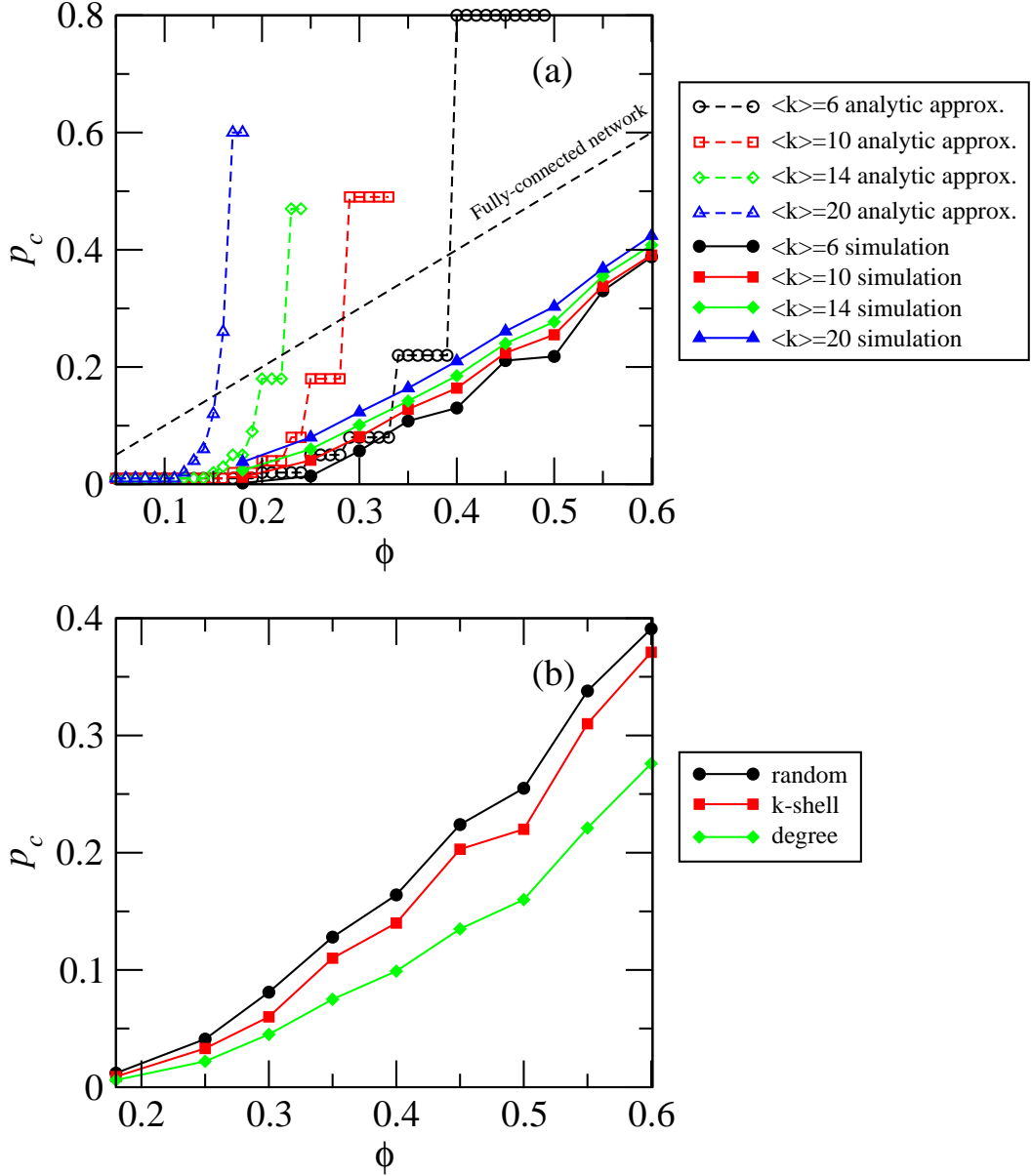}}
\caption{(a) Critical fraction of initiators (obtained by simulation and analytic approximation) for global cascades $p_{c}$
as a function of the local threshold-value $\phi$ for ER networks of
size $N=5000$ with various values of the average degree. The dashed
line corresponds to the exact limiting case on large complete graphs
(fully-connected networks), $p_{c}\simeq\phi$. (b) Critical fraction
of intiators for three different selection strategies for ER networks of $\langle k\rangle = 10$ and $N=5000$.}
\label{pc_phi}
\end{figure}

The critical point $p_{c}$ in each case is calculated by numerically
computing the derivative of $\tilde{S}$ with respect to $p$ and
finding its maximum. Having calculated $p_{c}$ allows us to
explicitly look at the relationship between $p_{c}$ and $\phi$ as
shown in Fig.~\ref{pc_phi}(a) for different average degrees $\langle
k\rangle$. As $\langle k\rangle$ increases, all curves appear to
converge to the limiting case of the fully-connected network
(complete graph) for which $p_{c}=\phi$. Therefore, for a given
threshold $\phi$ the minimum number of initiators needed to trigger
large cascades can be estimated.
We also employed a previously developed asymptotic
method~\cite{Gleeson_PRE2007} to estimate $p_{c}(\phi)$ analytically
(see Supplementary Information Section~S.3 for more details). This
method uses a tree-approximation for the network structure and
calculates the cascade size by assuming a progressive, directed
activation of nodes from the surface of the tree to the root.
Consequently, the method works well only for low $\langle k\rangle$
and low $p$. For large $\langle k\rangle$, the tree-approximation
breaks down, while for large $p$, deviations from the assumed
progressive and directed activation of levels, become significant.
The comparison of the analytically predicted $p_{c}$ using this
method to values obtained from simulations clearly show regions of
approximation validity and breakdown [Fig.~\ref{pc_phi}(a)].

For a fixed $\langle k\rangle=10$ and $N=5000$, we also studied by simulations how the
selection of initiators affect the critical fraction $p_{c}$.
Simulation results in Fig.~\ref{pc_phi}(b) show that selection of
initiators by their degree works better than the other two methods
across the range of threshold $\phi$.

\subsection*{Impact of network structure and clustering}

In this section we study how the dynamics of the threshold model is
affected by structural changes in the network. We study the dynamics
on an empirical high-school friendship network, using one particular
network from the Add Health data set (also employed in~\cite{qiminglu})
and a few degree-sequence preserving randomized
versions of it.
[{\it Add Health} was designed by J. Richard Udry, Peter S.
Bearman, and Kathleen Mullan Harris, and funded by a grant
P01-HD31921 from the National Institute of Child Health and Human
Development, with cooperative funding from 17 other agencies. For
data files contact Add Health, Carolina Population Center, 123 W.
Franklin Street, Chapel Hill, NC 27516-2524, addhealth@unc.edu,
http://www.cpc.unc.edu/projects/addhealth/ (Accessed June 20, 2013).]
To simplify things, we extract the giant component
from the high-school network which has $N=921$ nodes and $\langle
k\rangle \approx 5.96$. Hereafter, we only consider the giant component
of this network and refer to it as the high-school network. The initiator
fraction is kept fixed at $p=0.01$. The network contains two
communities which are roughly equal in size. We generate two
distinct ensembles of networks from this high-school network by employing the following randomization methods:
\begin{enumerate}
\item The link swap method (henceforth referred to as {\it x-swap}) in which two links are selected at random and
then one end point of a link is swapped with the end point of the
other link. An x-swap step is disallowed if it results in
fragmentation of the network. This swapping is done repeatedly so
that the network is randomized to an extent that any community
structure, local clustering, or degree-degree correlation is
eliminated~\cite{hgp,gionis2007,sharan05}.
\item The exact sampling method by Del Genio et al. (DKTB)~\cite{bassler2009},
a connected network is constructed from the degree sequence of the original network.
The algorithm takes as input the exact degree sequence of the network and joins the link stubs
from different nodes until every stub has been paired with another stub \cite{bassler2009,Kim_JPhysA2009}.
\end{enumerate}

\begin{figure}[H!]
\centerline{\includegraphics*[width=180mm]{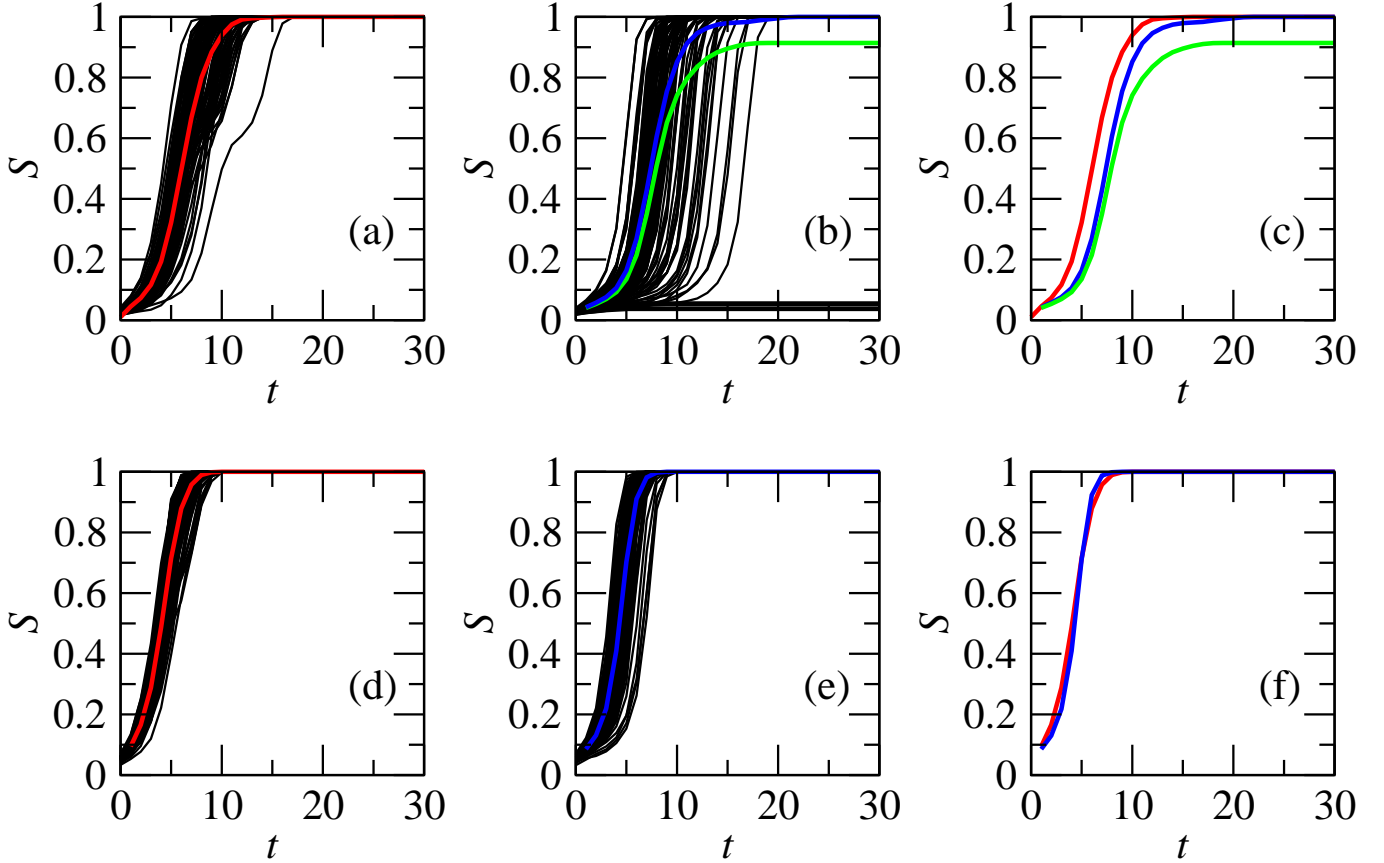}}
\caption{ Time
evolution of the size of the cascades $S$ on the high-school (HS)
network and its randomized version by X-swaps with identical degree
sequence, with $N=921$, $\langle k\rangle=5.96$, and $\phi=0.18$ for
two different values of fraction of initiators. (a) HS friendship
network and (b) its x-swapped randomized version.
(c) Direct comparison of the ensemble-averaged time series for the
original HS network (red solid curve) and for its x-swapped
randomized version (green solid curve); blue solid curves represent
conditional average over runs for which the spread reaches the
entire network. Thin black curves in (a) and (b) are individual time series. The fraction of initiators for (a--c) is $p=0.01$.
(d) HS friendship network and (e) its x-swapped randomized version.
(f) Direct comparison of the ensemble-averaged time
series for the original HS network (red solid curve) and for its
x-swapped randomized version; blue solid curves represent
conditional average over runs for which the spread reaches the
entire network. Thin black curves in (d) and (e) are individual time series. The fraction of initiators for (d--f) is $p=0.02$.
}
\label{s_t_xs_ex}
\end{figure}

\begin{figure}[H!]
\centerline{\includegraphics*[width=120mm]{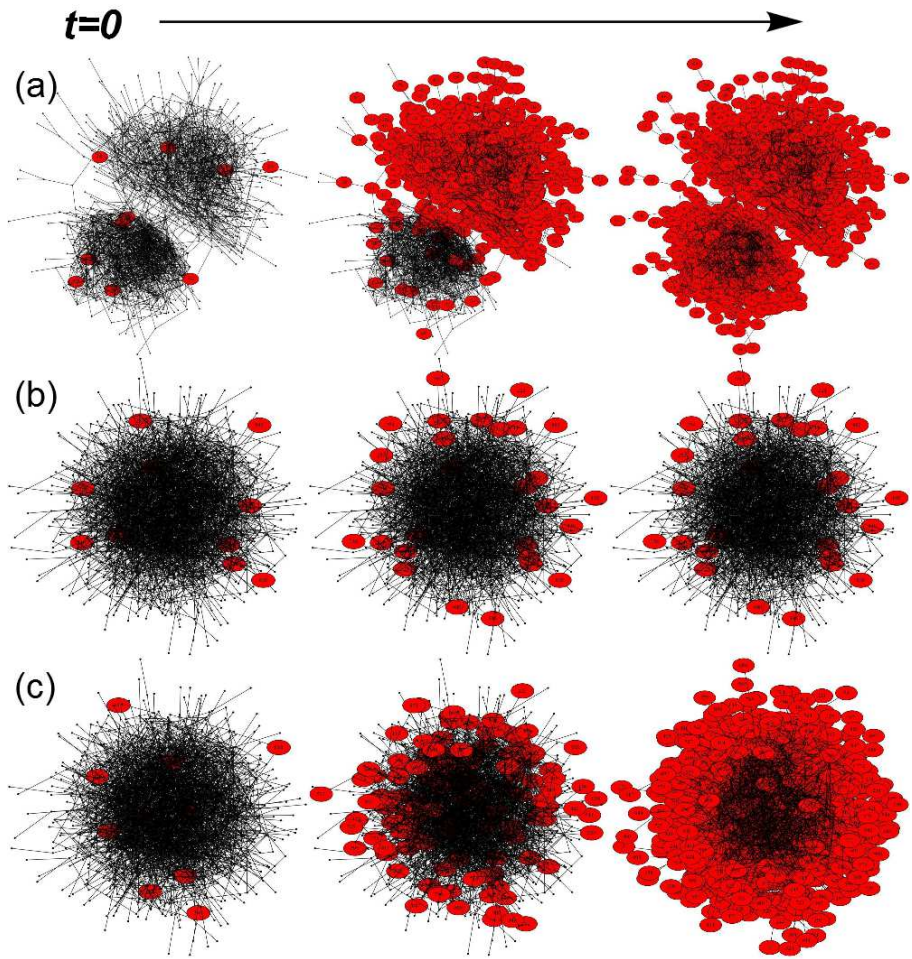} }
\caption{Visualizations of spreading in the threshold model (typical
individual runs) for various networks at different times during
evolution (arrow on top indicates the direction of time evolution).
$N=921$, $\langle k\rangle=5.96$, $p=0.01$ and $\phi=0.18$. Nodes in
state $1$ (active nodes) are colored red. (a) Original high-school
network; (b) Randomized network (by X-Swap) when eventual spread is
local; and (c) The same randomized network but for a run that
reaches the whole network.} \label{visuals}
\end{figure}

Both methods of randomization leave the degree sequence unchanged.
 (Results for x-swapped and exact sampling
\cite{bassler2009} are very similar and we only show them in detail
for the former.) We look at the size of spread $S$ as a function of
time for $p=0.01$ in the original high-school network
Fig.~\ref{s_t_xs_ex}(a) and the x-swapped high-school network
Fig.~\ref{s_t_xs_ex}(b), while Fig.~\ref{s_t_xs_ex}(c)
shows the direct comparison between the corresponding
ensemble-averaged time series. Analogous plots for $p=0.02$ are
shown in Figs.~\ref{s_t_xs_ex}(d--f).
For the empirical high-school network, some runs reveal the
existence of community structure in the network where spread is
faster in one community compared to other.
More specifically, in some of these runs, the cascade
first sweeps one of the communities (while the other one resists)
before it becomes global. This can be seen by the step-like
evolution in the corresponding time series in
Fig.~\ref{s_t_xs_ex}(a) [randomized networks do not exhibit this
behavior, see Figs.~\ref{s_t_xs_ex}(b)]. The same phenomena can also
be observed in the configuration snapshots in Fig.~\ref{visuals}(a),
while their randomized counterparts do not show this behavior
[Fig.~\ref{visuals}(b,c)].
In general, the results show that triggered cascades are larger and
more likely for a network with high local clustering than for a
randomized network with the same degree sequence
[Fig.~\ref{s_t_xs_ex}], although the impact of clustering is
diminishing for larger values of $p$. Note that the clustering
coefficient of the original high-school (HS) graph is
$C_{HS}\approx0.125$; for its randomized versions obtained by
x-swaps (XS) and exact-degree sequence (DKTB \cite{bassler2009})
construction are $C_{XS}\approx C_{DKTB}\approx0.008$ (see
Supplementary Information Section~S.4 for more details).

\begin{figure}[H!]
\centerline{\includegraphics*[width=130mm]{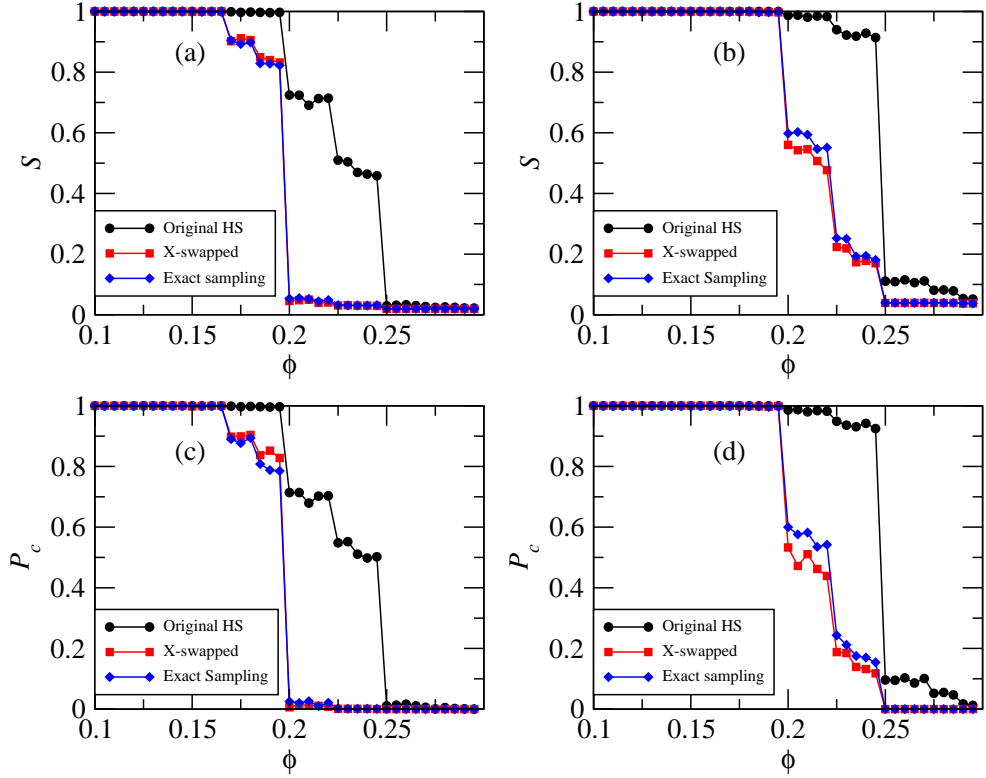}}
\caption{Cascade size and probability of global cascades in the high-school (HS) network and
its two randomized versions with identical degree sequence (X-swapped and exact sampling \cite{bassler2009})
with $N=921$ and $\langle k\rangle=5.96$.
Cascade size $S$ as a function of $\phi$ for (a) $p=0.01$ and for (b) $p=0.02$.
Probability of global cascades $P_{c}$ as a function of $\phi$ for (c) $p=0.01$ and for (d) $p=0.02$.}
\label{s_phi}
\end{figure}

The average cascade size $S$ [Fig.~\ref{s_phi}(a) and (b)] and the
probability of global cascades $P_{c}$ [Fig.~\ref{s_phi}(c) and (d)]
as a function of threshold $\phi$ also indicate that strong
clustering (present in empirical networks) facilitates
threshold-limited spreading. (We define a global cascade as a
cascade that covers at least $60\%$ percent of the network size
$N$.) Hence, this important feature of threshold-limited spreading
\cite{centola,Ikeda_JPCS2010} is preserved for the case of multiple
initiators studied here.

\begin{figure}[H!]
\centerline{\includegraphics*[width=120mm]{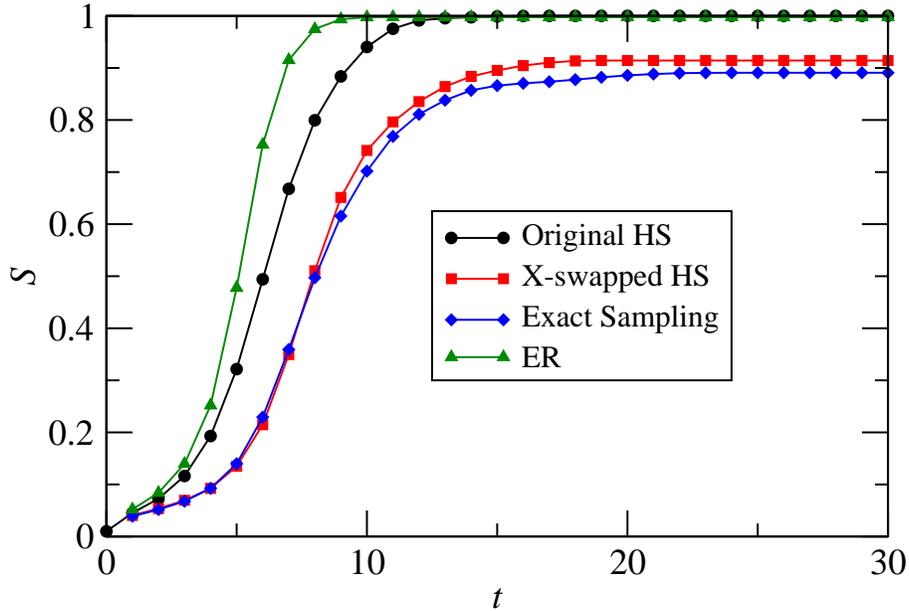}}
\caption{Average cascade size as a function of time in the high-school network (HS),
in its two randomized versions with identical degree sequence (x-swapped and exact sampling \cite{bassler2009}),
and in ER networks with the same average degree.
$N=921$, $\langle k\rangle=5.96$, $p=0.01$ and $\phi=0.18$ in all cases.}
\label{s_t_avg}
\end{figure}

The temporal evolution of the average cascade size in the original
HS network, its two randomized versions, and an ER network of the
same size and with the same average degree is shown in
Fig.~\ref{s_t_avg}. The two methods of randomization (x-swap and
exact sampling) roughly give the same cascade size $S$. In case of
randomized networks, for some realizations spread reaches the full
network [Fig.~\ref{visuals}(c)] and for some realizations spread is
minuscule [Fig.~\ref{visuals}(b)] and therefore $S<1$.

\begin{figure}[H!]
\centerline{\includegraphics*[width=160mm]{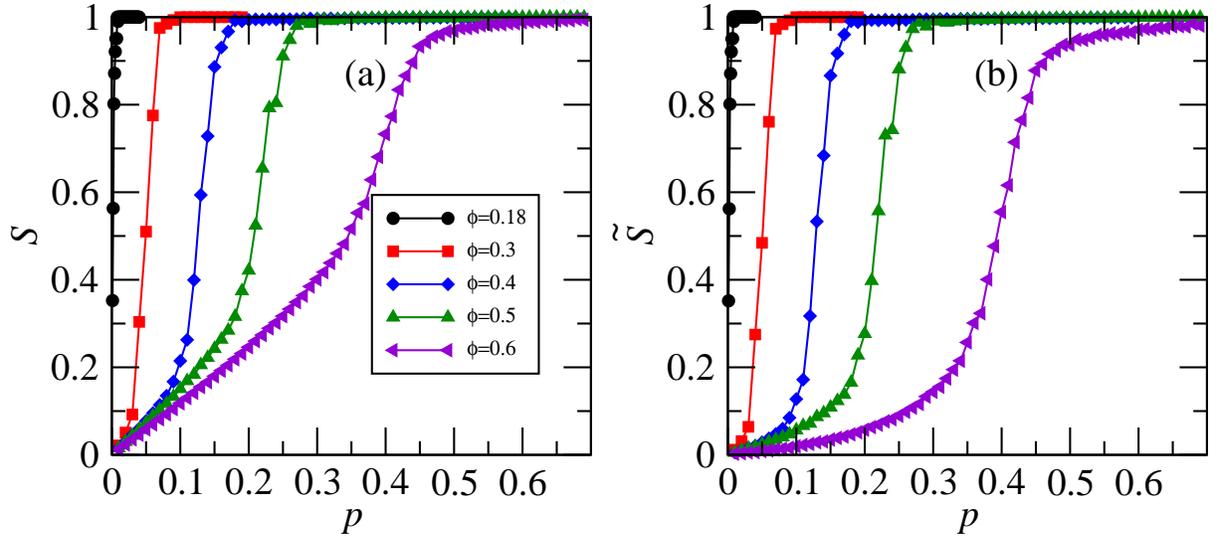}}
\caption{Cascade size and scaled cascade size as a function of initiators on the high-school network ($N=921$, $\langle k\rangle=5.96$).
(a) Cascade size $S$ as a function of initiators  $p$ for different values of $\phi$.
(b) Scaled cascade size $\tilde{S}$ [Eq.~(\ref{S_scaled})] vs. $p$ for different values of $\phi$.}
\vspace*{20.5truecm}
\label{s_p_hs}
\end{figure}

Finally, analogous to Fig.~\ref{sn_p_differentphis}, we show the
emergence of global cascades (at the tipping point $p_c$) in the
high-school network, as the density of initiators is varied
[Fig.~\ref{s_p_hs}].

\section*{Discussion}

Several recent studies have addressed, for a variety of agent-based
opinion spreading models, the impact of a special set of initiators
viz. inflexible individuals \cite{Galam_2007}, also referred to
synonymously as committed
\cite{qiminglu,xie2011,singh2012,Xie_PLOS2012,Zhang_PRE2012,Marvel_PRL2012,Halu_2012,Turalska_SREP2013}
or stubborn \cite{Yildiz_OR2011} agents, true believers
\cite{Centola_AJS2005}, zealots
\cite{Mobilia_PRL2003,Mobilia_JSAT2007,Swami_MILCOM2012}, or
inflexible contrarians \cite{Li_PRE2011,Li_JSP2012}. The rules of
state updating (or opinion switching) in these models is symmetric,
and governed purely by the local density of states in the
neighborhood of a node. In such a system, the inflexible nodes
constitute a special set of nodes which never change their opinion,
thereby breaking the symmetry of the system and giving rise to
tipping points beyond which the entire network conforms to the state
adopted by the committed agents. It has been shown that the
emergence of tipping points in some of these models is related to
metastable regions and barriers (saddle points) in the corresponding
opinion landscapes \cite{xie2011,Xie_PLOS2012,Zhang_PRE2012}.
Because these models allow frequent changes of state or opinion at
the individual level, these models are more suitable for scenarios
where switching an individual's state incurs virtually no cost.

In contrast to the above models, the threshold model (or the
qualitatively similar threshold contact process
\cite{Handjani_JTP1997,Dodds_PRL2004,Schonmann_PTRF2008,Schonmann_AP2009})
is more suited to modeling the diffusion of innovations or adoption
of new products where investment in a new idea comes at a cost, and
the incentive to switch back after becoming active is low. Here,
spreading is an asymmetric process and is also inhibited by a local
threshold: individuals can only adopt the new product or norm if a
sufficient fraction of their neighbors have already done so. (The
threshold model or threshold contact process, in spirit, is closer
to the family of Susceptible-Infected-Susceptible- or
contact-process-like models
\cite{kitsak,Harris_1974,Liggett_1999,Durrett_2010,Bass_MS1969,Chakrabarti_ACM2008,Prakash_WWW2012},
in that the spreading of a disease or norm is an inherently
asymmetric process by the rules of the local dynamics.)

The focus of this work was to identify tipping points for global
cascades triggered by multiple initiators and governed by local
thresholds. Our findings demonstrate that these tipping points emerge in
both ER and empirical high-school networks, in a qualitatively
similar fashion.

Further, we studied three different heuristic strategies to select a
fraction of initiators for the threshold model on ER network as well
as on an empirical network. Our results demonstrate that selecting
initiators by their degree (highest first) results in the largest
(as well as fastest) spread. Naturally, for high values of the local
threshold ($\phi>1/\langle k\rangle$), single initiators or small
cliques cannot trigger global cascades. We showed by simulations
that there exists a critical value of initiator fraction $p_c$ that
is needed to trigger cascades for high values of $\phi$. We also
studied how structural changes, such as randomizing an empirical
network using different randomizing methods, would affect the size
of the cascades triggered (in the cases studied here) by multiple
initiators. Our simulation results on the empirical high-school
network show that randomizing the network in fact results in
narrower cascade windows compared to the original network with
strong clustering, implying that clustering facilitates spreading in
threshold-limited diffusion with multiple initiators.

\section*{Acknowledgments}
This work was supported in part
by the Army Research Laboratory under Cooperative Agreement Number W911NF-09-2-0053,
by the Army Research Office grant W911NF-12-1-0546,
by the Office of Naval Research Grant No.~N00014-09-1-0607, and
by grant No.~FA9550-12-1-0405 from the U.S. Air Force Office of
Scientific Research (AFOSR) and the Defense Advanced Research Projects Agency (DARPA).
The views and conclusions contained in this document are those of
the authors and should not be interpreted as representing the
official policies either expressed or implied of the Army Research
Laboratory or the U.S. Government.

\section*{Author Contributions}
P.S., S.S., B.K.S. and G.K. designed the research;
P.S. and S.S. implemented and performed numerical experiments and simulations;
P.S., S.S., B.K.S. and G.K. analyzed data and discussed results;
P.S., S.S., B.K.S. and G.K. wrote, reviewed, and revised the manuscript.

\section*{Additional Information}
Competing financial interests: The authors declare no competing financial interests.


\vspace*{-10truecm}

\end{document}